\documentclass[sigconf]{acmart}

\usepackage{booktabs} % For formal tables
\usepackage{graphicx}
\usepackage{array}
\usepackage{caption}
\usepackage{algorithm}
\usepackage{algorithmicx} 
\usepackage{algpseudocode}
\usepackage{amsmath}  
\usepackage{graphicx}
\usepackage{multirow}
\usepackage{enumitem}
\usepackage{subfig}

\usepackage{array}
\newcommand{\PreserveBackslash}[1]{\let\temp=\\#1\let\\=\temp}
\newcolumntype{C}[1]{>{\PreserveBackslash\centering}m{#1}}
\newcolumntype{R}[1]{>{\PreserveBackslash\raggedleft}m{#1}}
\newcolumntype{L}[1]{>{\PreserveBackslash\raggedright}m{#1}} 

\makeatletter
\newcommand\fs@spaceruled{\def\@fs@cfont{\bfseries}\let\@fs@capt\floatc@ruled
  \def\@fs@pre{\hrule height.8pt depth0pt \kern2pt}%
  \def\@fs@post{\kern2pt\hrule\relax\vspace{-10pt}}%
  \def\@fs@mid{\kern2pt\hrule\kern2pt}%
  \let\@fs@iftopcapt\iftrue}
\makeatother

\setcopyright{acmcopyright}
\copyrightyear{2020} 
\acmYear{2020} 
\setcopyright{acmcopyright}
\setcopyright{acmcopyright}\acmConference[ICCAD '20]{IEEE/ACM International Conference on Computer-Aided Design}{November 2--5, 2020}{Virtual Event, USA \vspace{-2pt}}
\acmPrice{15.00 \vspace{-2pt}}
\acmDOI{10.1145/3400302.3415609}
\acmISBN{978-1-4503-8026-3/20/11}

\begin{document}
\title{DNNExplorer: A Framework for Modeling and Exploring a Novel Paradigm of FPGA-based DNN Accelerator}
%\titlenote{Produces the permission block, and
%  copyright information}
%\subtitle{Extended Abstract}
%\subtitlenote{The full version of the author's guide is available as
%  \texttt{acmart.pdf} document}

\author{\large Xiaofan Zhang$^{1*}$, Hanchen Ye$^{1*}$, Junsong Wang$^{2}$, Yonghua Lin$^{2}$, Jinjun Xiong$^{3}$, Wen-mei Hwu$^{1}$, Deming Chen$^{1}$ \\% <-this % stops a space 
        \large$^{1}$University of Illinois at Urbana-Champaign,
        $^{2}$Easy-visible, $^{3}$IBM Research\\
        \large \textit{\{xiaofan3, hanchen8, w-hwu, dchen\}@illinois.edu, \{junsong.wang, yonghua.lin\}@easy-visible.com,  jinjun@us.ibm.com}\vspace{-4pt}
}

% The default list of authors is too long for headers}
%\renewcommand{\shortauthors}{B. Trovato et al.}

\begin{abstract}
\vspace{-2pt}

Existing FPGA-based DNN accelerators typically fall into two design paradigms. Either they adopt a generic reusable architecture to support different DNN networks but leave some performance and efficiency on the table because of the sacrifice of design specificity. Or they apply a layer-wise tailor-made architecture to optimize layer-specific demands for computation and resources but loose the scalability of adaptation to a wide range of DNN networks. To overcome these drawbacks, this paper proposes a novel FPGA-based DNN accelerator design paradigm and its automation tool, called DNNExplorer, to enable fast exploration of various accelerator designs under the proposed paradigm and deliver optimized accelerator architectures for existing and emerging DNN networks. Three key techniques are essential for DNNExplorer's improved performance, better specificity, and scalability, including (1) a unique accelerator design paradigm with both high-dimensional design space support and fine-grained adjustability, (2) a dynamic design space to accommodate different combinations of DNN workloads and targeted FPGAs, and (3) a design space exploration (DSE) engine to generate optimized accelerator architectures following the proposed paradigm by simultaneously considering both FPGAs' computation and memory resources and DNN networks' layer-wise characteristics and overall complexity. Experimental results show that, for the same FPGAs, accelerators generated by DNNExplorer can deliver up to 4.2x higher performances (GOP/s) than the state-of-the-art layer-wise pipelined solutions generated by DNNBuilder \cite{zhang2018dnnbuilder} for VGG-like DNN with 38 CONV layers. Compared to accelerators with generic reusable computation units, DNNExplorer achieves up to 2.0x and 4.4x DSP efficiency improvement than a recently published accelerator design from academia (HybridDNN \cite{ye2020hybrid}) and a commercial DNN accelerator IP (Xilinx DPU \cite{xilinx_dpu}), respectively.

%\vspace{-15pt}
\end{abstract}

% \keywords{\vspace{-4pt}Automation Design Flow, FPGAs, DNNs, Design Space Exploration\vspace{-12pt}}

\maketitle
\vspace{-10pt}
\section{Introduction}
\vspace{-2pt}
%Outline
%1)intro advantages of using FPGA
%2)intro the two paradigms
%3)intro the challenges/drawbacks of using these paradigms, and how we can help
%4)summary the novelty of the paper
% \quad 
Great successes have been achieved by deep neural networks (DNNs) in a massive number of real-life artificial intelligence (AI) applications. Such a success has been driven in part by the continuous improvement of DNN models, using deeper and more sophisticated
layer interconnections, to deliver the state-of-the-art solutions \cite{krizhevsky2012imagenet,simonyan2014very,szegedy2015going,he2016deep,redmon2016you,zoph2018learning,real2019regularized,zhang2020skynet}. With the impressive quality of results comes along DNN models' increasing requirements of more computation and memory. This in turn puts stringent design constraints on any domain-specific hardware accelerator designs to not only deliver high inference accuracy but also satisfy inference speed, throughput, and energy efficiency. 

There are rich studies on customized DNN accelerators that take advantage of different hardware devices, such as exploring kernel optimizations on GPUs and customizing accelerators on ASICs and FPGAs
%to provide improved speed and efficiency of DNN inference and training 
\cite{franklin2018nvidia,jouppi2017datacenter,isscc_2016_chen_eyeriss,zhang2015optimizing,xu2020autodnnchip}. Because of FPGA's flexibility of meeting stringent energy constraints and fast  time-to-market, FPGA-based solutions have recently become prominent choices for DNN acceleration with improved latency and energy efficiency
%compared to GPUs while offering much more flexibility than ASICs 
\cite{qiu2016going,zhang2017high,zhang2017machine}.

%2) intro the two paradigms
By investigating recent FPGA-based solutions, we see two popular accelerator design paradigms: one adopts the generic reusable architectures for all DNN layers; while the other customizes
hardware implementation for each DNN layer. When following the first  paradigm, such as designs in \cite{zhang2017improving,hao2019fpga,ye2020hybrid}, a generic computation unit is instantiated on FPGA where all DNN layers are processed in a recurrent manner.
% , which is effective to adapt regular and deep networks. 
When adopting the second paradigm, such as designs in \cite{li2016high,zhang2018dnnbuilder,wei2018tgpa}, a layer-wise pipeline architecture is implemented on FPGA, where each DNN layer is handled by a dedicated pipeline stage. In other words, accelerator designs have better customization for each layer.
%
%3) intro the challenges/drawbacks of using these paradigms, and how we can help
However, both paradigms may not be suitable for the emerging DNN models where more diverse layer configurations and deeper network structures are involved. The first paradigm has difficulty to optimize DNNs with diverse layers
% large inputs, as the increasing amount of intermediate results (feature maps) can easily exhaust the scarce FPGA on-chip memory and force to use external memory. The frequent data swap  between on-chip and external memory also occupies external data access bandwidth originally assigned for loading DNN parameters. In addition, the diverse layers 
that exhibit vastly different computation-to-communication (CTC) ratios. This will degrade the accelerator performance because of the lack of fine-grained adjustments in such a generic architecture. The second paradigm also has an obvious flaw as it requires hardware instances for each pipeline stage. The more layers in the DNN models,
% , the more stages need to be instantiated. This 
the less resources for each stage, which eventually leads to  lower performance.
% In pipelined designs, any lacks of resources will become a bottleneck.We will provide more in-depth discussions in Sec. XX. 

To address these challenges, we propose DNNExplorer, a framework to model and explore DNN accelerator designs with the goal of producing a balanced DNN  solution while keeping the advantages of the two aforementioned popular accelerator design paradigms by addressing their respective flaws.
% and deliver optimized performance. 
In summary, the main contributions of this paper are as follows.

\begin{enumerate}[itemindent=0em,fullwidth]
% \vspace{-2pt}

\item We introduce a new FPGA-based accelerator design paradigm with higher-dimensional design space support, better specificity, and improved scalability to effectively address the drawbacks of existing FPGA-based DNN accelerator designs.

\item We propose an automation tool called DNNExplorer to enable fast architecture exploration under the proposed paradigm and deliver optimized accelerator designs in three steps as: \textit{model analysis}, \textit{performance modeling}, and \textit{architecture exploration}.
 
\item We define a dynamic design space to capture more detailed configurations of the proposed accelerator paradigm. It helps generate a well-defined design space for architecture exploration to better accommodate different input DNN models and targeted FPGAs. 

\item We propose a two-level (global and local) automatic design space exploration (DSE) engine to efficiently find optimized accelerator architectures by simultaneously considering both FPGAs' computation and memory resources and DNN networks' layer-wise characteristics and overall complexity. 

\item We experimentally demonstrate that DNNExplorer delivers better FPGA-based DNN accelerators with 4.2x higher throughput compared to the pure pipeline design (DNNBuilder), and 4.4x higher DSP efficiency compared to a generic design (Xilinx DPU). 
\vspace{-2pt}
\end{enumerate}

%% paper structure
% The rest of the paper is organized as follows. Sec. 2 introduces the related work while Sec. 3 summarizes the main challenges we encountered. Sec. 4 and 5 describe the design flow of DNNBuilder and the architecture of generated accelerators. The resource allocation scheme is presented in Sec. 6. Sec. 7 shows the experimental results and Sec. 8 concludes this paper.
% \vspace{-2pt}
\vspace{-5pt}
\section{Related Work} 
\vspace{-2pt}
%% FPGA accelerator

There are extensive studies in customized DNN accelerator design using FPGAs. To pursue higher energy efficiency, accelerators in \cite{qiu2016going,zhang2017high} incorporate a dynamic data quantization scheme for both DNN parameters and intermediate feature maps to relax the compute and memory demands. The designs in \cite{zhao2017accelerating,wang2018design} support binary and ternary quantization, which intend to replace the hardware-intensive floating-point multiplications by logical operations. 
Fast convolution (CONV) algorithms have been investigated to lower the computation complexity and speedup DNN inference, such as using Winograd-based solutions, fast Fourier transform (FFT), and hybrid schemes with both spatial CONV and fast CONV algorithms \cite{xiao2017exploring, zhuge2018face,ye2020hybrid}.
A fine-grained layer-based pipeline architecture is proposed by DNNBuilder\cite{zhang2018dnnbuilder}, which works with a resource allocation scheme to lower the end-to-end latency when running real-life DNN applications. 
Recently, more solutions have been developed with both hardware and software optimizations. For example, authors in \cite{han2017ese} apply DNN compression before mapping the DNNs onto FPGAs for higher speedup, while authors in \cite{hao2019fpga, zhang2020skynet} propose hardware-software co-design  to accelerate resource-demanding DNN applications for embedded FPGAs.

%% accelerator framework
Recently published literature also focuses on building automation tools for rapid performance estimation and implementation of FPGA-based DNN accelerators. 
The work in \cite{zhang2018caffeine} proposes a unified representation for CONV and fully-connected (FC) layers to facilitate the accelerator modeling on targeted FPGAs. A framework proposed in \cite{wei2017automated} incorporates systolic arrays to improve computing efficiency, while the design in \cite{ye2020hybrid} further improves the accelerator design by proposing processing engines with both spatial- and Winograd-CONV support. To help fully explore available resources of the targeted FPGAs, authors in \cite{zhang2018dnnbuilder} introduce an automatic resource allocation tool to generate fine-grained parallelism guidelines for its architecture. 
%% HLS
Researchers also employ High-level Synthesis (HLS) in their tools to improve design efficiency of FPGA-based hardware designs \cite{rupnow2011study,liu2016high,guan2017fp,li2019implementing,chen2019cloud}.

\begin{figure}[t]
  \centering
  \vspace{-15pt}
  \includegraphics[width=1.05\columnwidth]{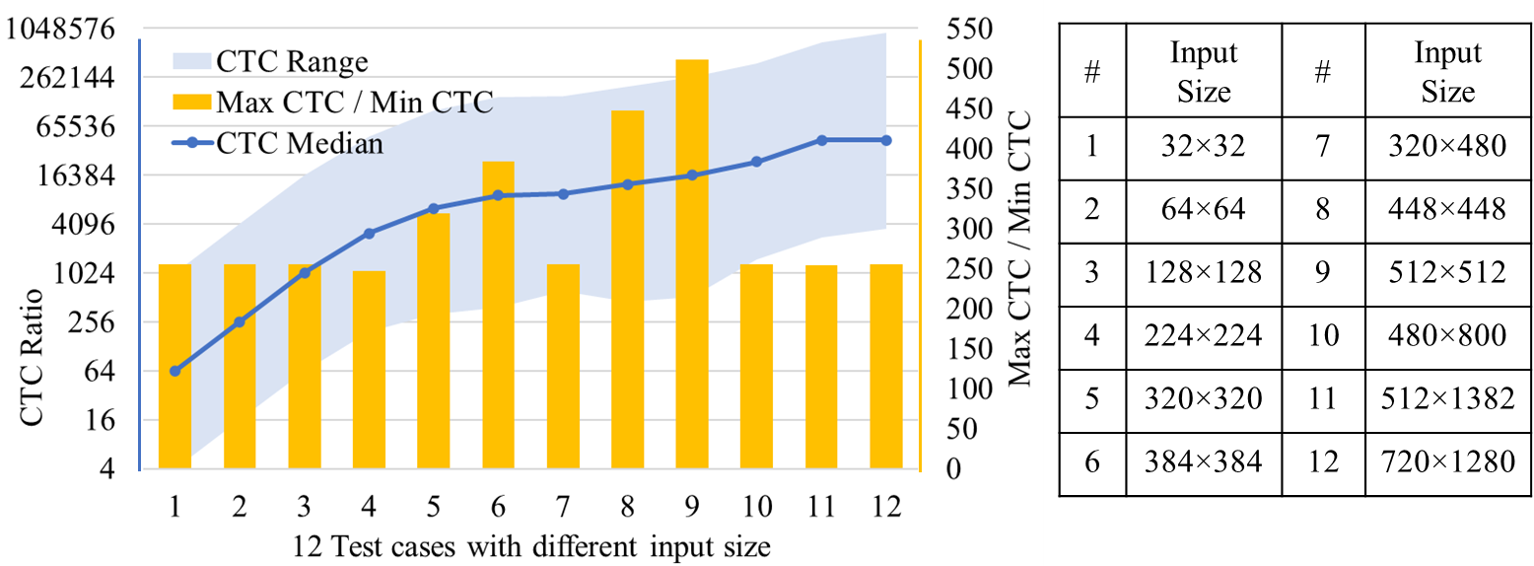}
   \vspace{-15pt}
   \caption{CTC (computation-to-communication) distribution in VGG-16 models (without FC layers) regarding 12 input resolution cases. Inputs are RGB images with 3 depth channels and height and width listed as input size.}
   \label{fig:inputsize_ctc}
   \vspace{-20pt}
\end{figure}

\vspace{-5pt}
\section{Challenges of current solutions}
\vspace{-2pt}
%intro the existing paradigms
% highlight draw for acc --> result CTC VS perf
% highlight drawback for pip --> result perf VS depth

\begin{figure}[t]
  \centering
  \vspace{-20pt}
  \includegraphics[width=1.05\columnwidth]{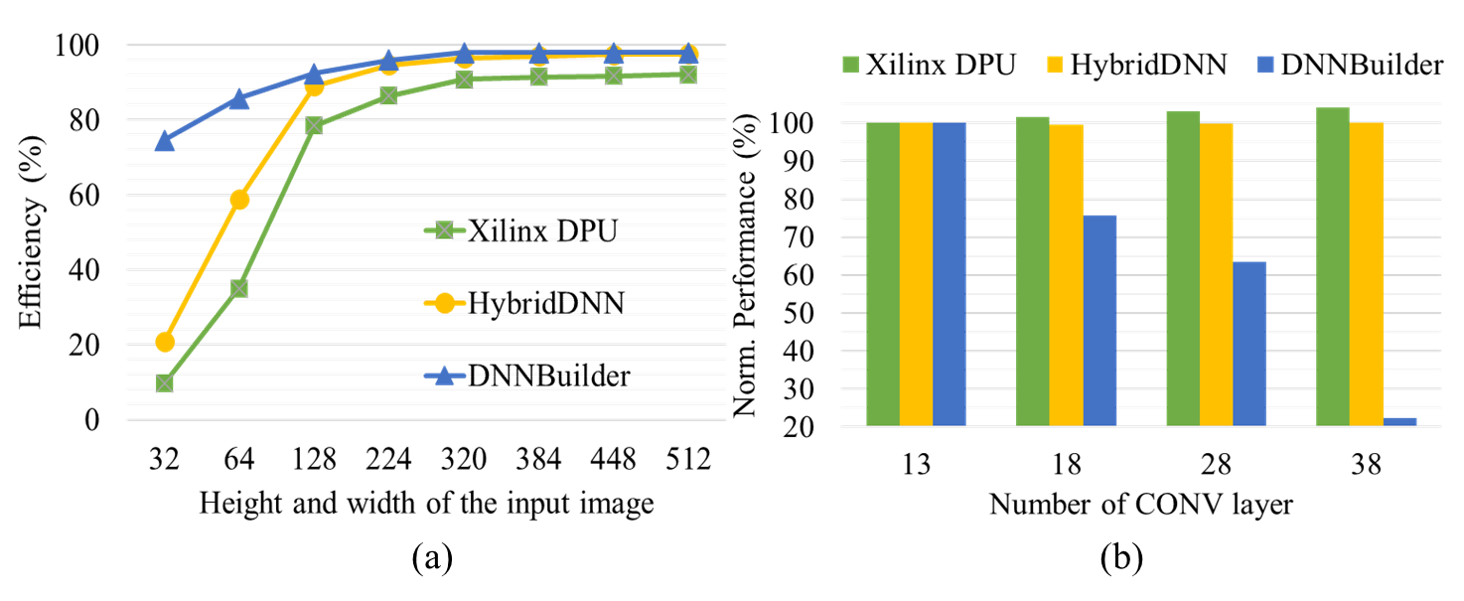}
   \vspace{-25pt}
   \caption{(a) The trends of DSP efficiency when running VGG16 with increasing input sizes in three representative FPGA-based DNN accelerators (batch size $=$ 1); (b) Normalized throughput performance in three accelerators when running VGG-like DNNs with 3$\times$224$\times$224 inputs and 13$\sim$38 CONV layers. (The performance of each accelerator is normalized to their baseline cases running the 13-layer DNN.)}
   \label{fig:inputsize_depth}
   \vspace{-12pt}
\end{figure}

% Xilinx DPU and HybridDNN follow the first design paradigm while DNNBuilder uses the second one;
% As discussed in Section 1, previous FPGA-based DNN accelerator designs may suffer several drawbacks. 
By following the first paradigm as mentioned in Section 1, different DNN layers are handled by a generic compute unit. Although the majority of computation in DNNs comes from CONV layers, the differences among CONV layers may be significant. For example, Figure \ref{fig:inputsize_ctc} shows 12 models' layer characteristics in terms of their CTC ratio distribution.
%highlight some num
The medians of CTC show an upward trend along with higher input resolutions. From 32$\times$32 to 512$\times$512 inputs, CTC medians rapidly increase by nearly 256 times following the blue curve, which implies 
% means the same DNN model with different input resolutions can 
 significantly different computation and memory-access patterns. 
It is also obvious that the CTC range or variations are large even for the same model.
% More prominently, the CTC variances are even greater between layers in the same model. Shown by the yellow bars, the upper-bound of the CTC can be 500x higher than the lower bound in case \#9. 
If using a generic compute unit to process these layers, we have to accept sub-optimal solutions because of the lack of architecture specificity and limited design spaces. 
We use DSP efficiency (Eq. \ref{eq:dsp_eff}) to evaluate whether an accelerator is working efficiently, where $\alpha$ represents the number of multiply-accumulate (MAC) operations handled by one DSP in one clock cycle, i.e.,  $\alpha=2$ for 16-bit and $\alpha=4$ for 8-bit inputs.
\begin{equation}
\small
\label{eq:dsp_eff}
    EFFI_{DSP} = \frac{GOP/s}{\alpha\times DSP_{allocated}\times FREQ}
\end{equation}

As shown in Figure \ref{fig:inputsize_depth} (a), both Xilinx DPU \cite{xilinx_dpu} and HybridDNN \cite{ye2020hybrid} (representing the first paradigm) suffer lower DSP efficiency (up to 64.9\% and 53.7\% degradation, respectively) compared to dedicated designs from DNNBuilder \cite{zhang2018dnnbuilder} (representing the second paradigm).
For accelerators following the second paradigm, separate pipeline stages are instantiated on FPGAs for each major DNN layer and eventually all of them are combined into pipeline implementations. This allows dedicate layer designs according to layers' inherent characteristics (e.g., CTC ratio, computation and memory demands), which enables the possibility of more fine-grained hardware configurations and more in-depth design space exploration. However, its scalability may be easily restricted by the number of DNN layers that it can support, as a deeper DNN means more pipeline stages and less resources for each layer. In this case, performance degradation is expected. Some intuitive examples can be found in Figure \ref{fig:inputsize_depth} (b), where accelerators need to handle 4 VGG-like DNNs with 13$\sim$38 CONV layers. The performance of DNNBuilder decreases 77.8\%  on a 38-layer model compared to the shallower network with 13 CONV layers. In contrast, the generic accelerators maintain a stable performance.

% what we are going to do
% Our proposed tool can well address these drawbacks and provide a built-in design space exploration tool for high-performance and efficient designs.
\vspace{-5pt}
\section{DNNExplorer Framework}
\vspace{-2pt}

\vspace{-1pt}
\subsection{The Proposed Novel Paradigm}
\begin{figure}[t]
  \centering
  \vspace{-20pt}
  \includegraphics[width=0.9\columnwidth]{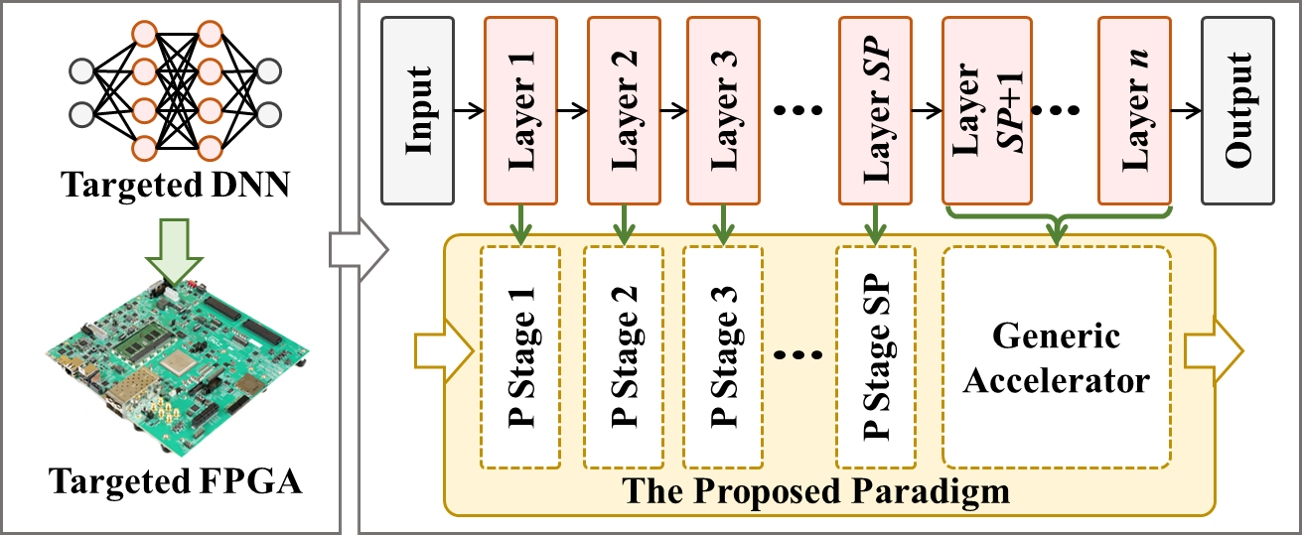}
   \vspace{-10pt}
   \caption{The proposed accelerator paradigm to overcome existing drawbacks when mapping DNNs onto FPGAs. It includes $SP$ pipelined stages for layer $1\sim SP$ and a generic accelerator for the rest of layers. $SP$ is the split-point.}
   \label{fig:new_paradigam}
   \vspace{-8pt}
\end{figure}

\vspace{-2pt}
%% intro the paradigm (pip+acc)
To overcome these challenges, we propose a new accelerator paradigm to capture the essences of both existing paradigms but address their disadvantages. We present the proposed paradigm in Figure \ref{fig:new_paradigam}, to leverage both layer-dedicated and generic reusable designs. For the first part, we implement a pipeline accelerator to generate dedicate layer stages for the first $SP$ layers. It helps guarantee higher DSP efficiency and more fine-grained resource allocation. For the second part, we adopt a generic architecture for the rest of DNN layers so that it supports deeper DNNs with given resources.

To validate the proposed idea, we investigate 10 popular DNNs and summarize their CTC fluctuations in Table \ref{tab:ctc_var}. 
% Variances of the CTC are used here as indicators. 
Specifically, we divide every DNN into two halves: the first half covers the bottom part of layers (close to the input layer) with 50\% of the total MAC operations while the second half contains the rest of layers. 
In each half, we calculate the average value of the squared difference between each layer's CTC and the mean CTC. We use  
$V_1$ and $V_2$ to represent these results (variances) in the first and second half, respectively. 
In Table \ref{tab:ctc_var}, $V_1$ is on average 1806.2 times higher than $V_2$, which means the first half have more CTC fluctuations. 
% compared to the second half. 
This is a common phenomenon during DNN inference, which causes low DSP efficiency of the FPGA-based accelerators following the first paradigm. %% Why the proposed one can fix ACC drawbacks
In our proposed design, these CTC fluctuations from the first half can be properly resolved by the layer-dedicated pipeline architecture. 
%% Why the proposed one can fix pip drawbacks
By considering the second half, where layers have less CTC variances and share higher similarity, a generic structure can successfully eliminate the poor-scalability of the pipeline design.

%% intro the dim in RSV (split-point, batch, DSP, BRAM, BW)
The proposed design is not a simple concatenation of the two existing paradigms,
% . Although it can potentially address the existing problems, 
as the fusion of two heterogeneous structures can directly cause an exponential increase of the design space and easily lead to tedious explorations and sub-optimal solutions. 
These additional design parameters include the task partitioning scheme, the batch size, and the hardware resource allocation between two structures. 
Therefore, we propose a highly-efficient design tool, called DNNExplorer, to perform automatic architecture exploration under the proposed paradigm and deliver optimized solutions.

\begin{table}[t]
\footnotesize
%\scriptsize
%\vspace{-5pt}
\caption{Ratio of CTC variances between the first half ($V_1$) and the second half ($V_2$) of DNNs.}
\vspace{-12pt}
\label{tab:ctc_var}
\begin{center}
\newcommand{\tabincell}[2]{\begin{tabular}{@{}#1@{}}#2\end{tabular}}
\begin{tabular}{C{0.06\textwidth}|C{0.07\textwidth}|C{0.05\textwidth}
|C{0.07\textwidth}|C{0.07\textwidth}|C{0.05\textwidth}}
\toprule
\textbf{Network} & \textbf{Input Size} & \textbf{$V_1 / V_2$} & \textbf{Network} & \textbf{Input Size} & \textbf{$V_1 / V_2$}\\ \hline
Alexnet & 3x227x227  & 185.8 & ResNet-18 & 3x224x224 &  1607.3 \\ \hline
GoogLeNet & 3x224x224  & 3622.8 & ResNet-50 & 3x224x224 & 998.7\\ \hline
InceptionV3 & 3x299x299 & 6210.6 & SqueezeNet & 3x227x227 & 238.9   \\ \hline
VGG-16 & 3x224x224 & 489.8  & MobileNet & 3x224x224 & 3904.2  \\ \hline
VGG-19 & 3x224x224 &  552.6 & MobileNetV2 & 3x224x224 & 251.5
  \\
\bottomrule
\end{tabular}
\vspace{-10pt}
\end{center}
\end{table}

\begin{figure}[t]
    \vspace{-20pt}
    \centering
    \includegraphics[width=1.05\columnwidth]{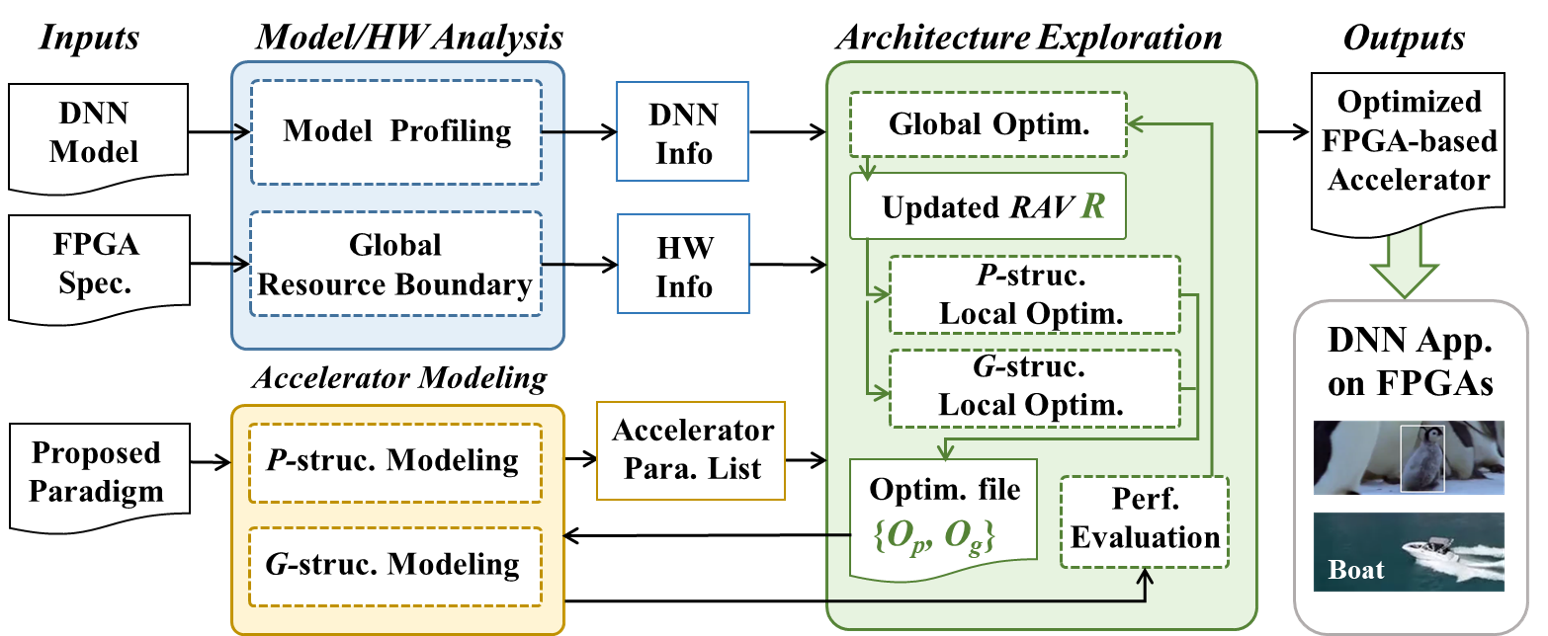}
    \vspace{-20pt}
    \caption{The Design flow of DNNExplorer containing 3 steps 
%   as (1) Model/Hardware Analysis, (2) Accelerator Modeling, and (3) Architecture exploration, 
   to generate optimized DNN accelerators given DNN models and FPGA specifications.}
   \label{fig:flow}
   \vspace{-20pt}
\end{figure}

\vspace{-5pt}
\subsection{DNNExplorer Design Flow}
\vspace{-2pt}

DNNExplorer generates optimized FPGA-based accelerators according to the input DNN model, the targeted FPGA specification, and the proposed accelerator paradigm. As shown in Figure \ref{fig:flow}, three steps are included in DNNExplorer to provide \textit{Model/HW Analysis}, \textit{Accelerator Modeling}, and \textit{Architecture Exploration}. 

In \textit{Model/HW Analysis}, 
% a targeted DNN and a FPGA specification are passed to DNNExplorer as inputs. 
% In general, the DNNs are developed, trained and fine-tuned by deep learning frameworks (e.g., Caffe\cite{jia2014caffe}, Tensorflow\cite{abadi2016tensorflow}, and PyTorch\cite{paszke2019pytorch}). After training for specific applications, 
DNN definition files and trained parameters are passed to DNNExplorer for model profiling. The layer-wise information
%, which distinguishes the input DNN, 
is extracted in this step, such as layer type, layer configuration, computation and memory demands, CTC ratio, quantization scheme, etc, and then packed as DNN info for \textit{Architecture Exploration}. The input also includes a FPGA specification, which helps setup boundaries of available resources, such as DSP, BRAM, and external memory bandwidth. %($DSP, BRAM, BW$). 
% In addition, initial hardware resource allocation is performed here. We use Resource Allocation Vector (RAV), a pair of 5-dim vectors \{$R_1,R_2$\}, to describe the resource allocated to the first (pipelined) and second (generic) structures of the proposed paradigm. In each vector, it contains the split-point ($SP$), which indicates task partitioning between the first and second structures, the batch size ($Batch$), and three major FPGA computation/memory resources as DSP, BRAM, and external access bandwidth ($DSP, BRAM, BW$). The initial RAV is then passed to \textit{Architecture Exploration} for two-level (global and local) optimizations.

In \textit{Accelerator Modeling}, we use models corresponding to the pipeline ($P$) and generic ($G$) structure of the proposed paradigm. The goal of this step is to adopt  highly-accurate pre-built analytical models for resource utilization and performance estimation. Eventually, a list of configurable accelerator parameters are collected for exploration in the next step. Also, the analytical models contribute to the performance evaluation in \textit{Architecture Exploration}.
In-depth introductions of these models are provided in Section 6.

In \textit{Architecture Exploration}, we adopt a divide and conquer strategy and design a two-level automatic DSE engine to efficiently search for optimal accelerators in the massive design space. 
We use Resource Allocation Vector (RAV), 
% a pair of 5-dim vectors \{$R_p,R_g$\}, 
a 5-dim vector $R$, 
to describe resources allocated to the $P$ and $G$ structure. $R$ contains the split-point ($SP$), which indicates task partitioning between these two structures, the batch size ($Batch$), and three major FPGA resources ($DSP, BRAM, BW$).
Given inputs from the last two steps, a global optimization is first performed to explore and update $R$ in every iteration, providing guidelines for building $P$ and $G$ structure, respectively. After that, local optimizations are executed individually for $P$ and $G$ structure to explore the best configurations. 
All selected accelerator parameters are documented on an optimization file for driving the performance evaluation using analytical models in the modeling step. With the performance feedback, the global optimization algorithm can decide whether to continue exploration. Eventually, DNNExplorer generates the optimized architecture.
% among the massive design space.

\vspace{-5pt}
\section{Accelerator Architecture}
\vspace{-2pt}

\vspace{-1pt}
\subsection{Accelerator Overview}
\vspace{-2pt}

In our design, the targeted DNN is mapped into hardware based on the RAV. Given a specific $R$, we have
\vspace{-5pt}
\begin{equation}
\small
R=[SP,Batch,DSP_p,BRAM_p,BW_p]
\end{equation}
% \begin{equation}
% \small
% R_g=[SP,Batch,DSP_g,BRAM_g,BW_g]
% \end{equation}

Layer $1 \sim$ $SP$ are instantiated into a layer-wise tailor-made structure following a pipeline manner with batch size $=Batch$ and available resource $[DSP_p,BRAM_p,BW_p]$. Denoting the total available global resource  on the targeted FPGA is $[DSP, BRAM, BW]$, the rest of layers are mapped into a generic reusable structure following a recurrent manner with batch size $=Batch$ and available resource $[DSP-DSP_p,BRAM-BRAM_p,BW-BW_p]$. The pipeline and generic structure share the same clock frequency. We will discuss the selection of RAV in Section 7.

\begin{figure}
    \vspace{-5pt}
    \centering
    \includegraphics[width=0.4\textwidth]{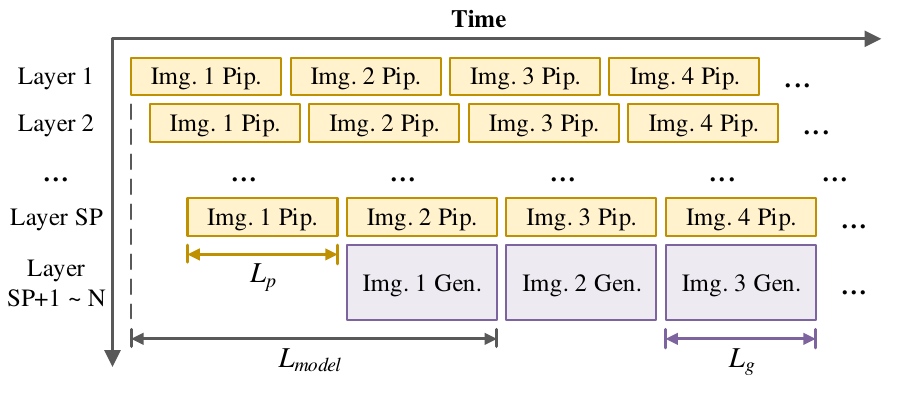}
    \vspace{-15pt}
    \caption{The overall dataflow of the proposed accelerator }
    \label{fig:dataflow}
    \vspace{-15pt}
\end{figure}

\begin{figure*}
    \vspace{-20pt}
    \centering
    \includegraphics[width=0.85\textwidth]{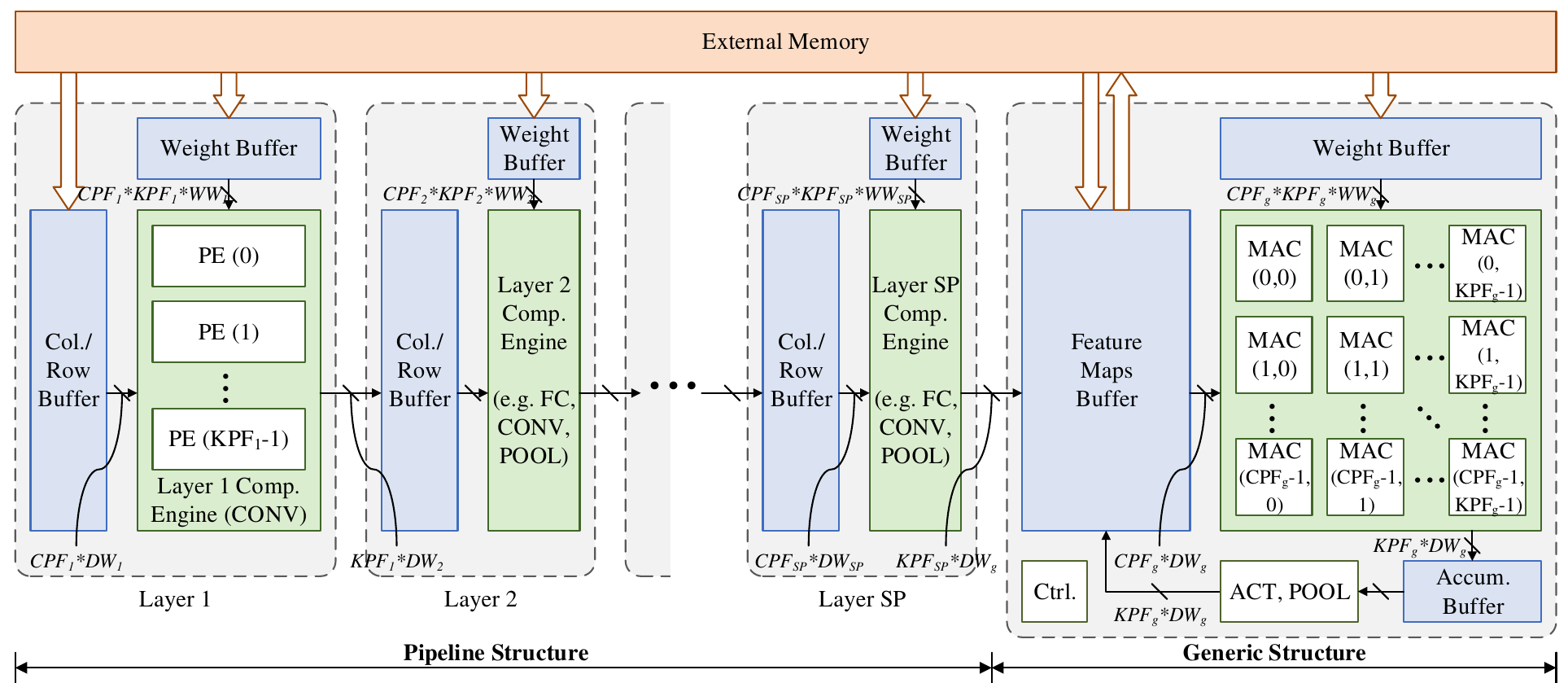}
    \vspace{-10pt}
    \caption{The proposed accelerator architecture with $SP$ pipeline stages instantiated for processing the first $SP$ DNN layers and a generic computation unit implemented for accelerating the rest of layers.}
    \label{fig:big_arch}
    \vspace{-10pt}
\end{figure*}

%
% Presuming the targeted DNN model is composed of $N$ layers, a tunable parameter $N_{split}$ (split point) is defined in our novel paradigm for indicating the workload partition for the targeted DNN model. Layer $1$ to layer $N_{split}$ are mapped to the first structure, while layer $N_{split}+1$ to layer $N$ are folded and processed by the second structure. $N_{split}$ is initialized and optimized in the phase of DSE, which will be discussed in Section 6.
%
Figure \ref{fig:big_arch} presents the proposed accelerator design, where three major FPGA resources are utilized as computation resources (green area), on-chip memory (blue area), and external memory (orange area). 
% We use three grey dash-line modules for representing the first part of  tailor-made layer instances implemented on FPGA, and one green dash-line modules for the second structure which will process the last $N-N_{split}$ folded layers. 
Inputs are first streamed into on-chip buffer from external memory in the first pipeline stage and then intermediate results (DNN feature maps) are passed horizontally through all pipeline stages and reach the feature map buffer of the generic structure. 
% As the same computation unit is shared by different layers, intermediate results need to swap in and out between on-chip and external memory once the computations of current layer are completed. 
On the other hand, trained DNN parameters are passed vertically with read-only transmission from the external memory. 
The overall dataflow of the proposed accelerator is shown in Figure \ref{fig:dataflow}. We adopt the fine-grained pipeline from \cite{zhang2018dnnbuilder} to reduce the initial latency when handling layer $1 \sim SP$ in our pipeline structure. After that, the generic structure starts working on the rest of layers. To reach the maximum throughput performance, we need to balance the latency of each pipeline stage ($L_p$) and the generic structure ($L_g$). With a perfect load-balanced design, we are able to achieve the peak throughput performance as $1/max(L_p,L_g)$. The detailed algorithms of searching such load-balanced design are introduced in Section 7.

\vspace{-5pt}
\subsection{The Pipeline Structure}
\vspace{-2pt}

The first part of Figure \ref{fig:big_arch} covers $SP$ dedicated pipeline stages to process major layers in targeted DNNs, such as convolutional (CONV) layers, pooling (POOL) layers, and fully-connected (FC) layers. These layers are likely to dominate most of the computation and memory resources. Other layers, such as batch normalization (BN) and activation layers are concatenated into the major ones and processed by the same pipeline stage.    
In each stage, we implement three main components as: (1) a computation engine to carry out DNN layer operations with dedicated parallelism design, (2) a weight buffer to pump in DNN parameters from external memory; and (3) a column/row buffer to cache columns or rows of intermediate results generated by the previous stage.
Four configurable parameters are available in every stage, including the channel parallelism factor (CPF) and the kernel parallelism factor (KPF) (which are unrolling factors along input and output dimensions), the input data bit-width (DW), and the weight bit-width (WW). With these parameters, we can generate dedicate accelerator designs for different DNN layers.

\subsubsection{\textbf{Computation Engine (CE)}} 
As shown in the green parts of Figure \ref{fig:big_arch}, CEs handle most of the computations from DNN layers. Inside each CE, we implement a number of processing elements (PEs) with two-dim parallelism (CPF and KPF). Since both CONV and FC layers share similar computation patterns, we leverage the same PE for these two layers. More specifically, assuming layer stage 1 in Figure \ref{fig:big_arch} works for CONV layers, the CE in this stage contains $KPF_1$ PEs, and each PE is designed to handle one $CPF_1$-length vector multiplication in one clock cycle (assuming $DW_1=WW_1=16$). 
% $CPF$ and $KPF$ is the abbreviation of Channel Parallel Factor and Kernel Parallel Factor respectively, which means we are unrolling dimensions of input and output channels in the CONV and FC computation engine. 
Once the trained DNN parameters are ready in weight buffer, we will broadcast a $CPF_1$-length vector of input feature map to all $KPF_1$ PEs, and after calculations, the accumulated results ($KPF_1$-length vector) will be written to the column/row buffer of the next stage.

% \begin{figure}[t]
%     \centering
%     \includegraphics[width=0.45\textwidth]{}
%     \caption{Pipeline Column-based Buffer}
%     \label{fig:pipeline_time}
% \end{figure}

\subsubsection{\textbf{Fine-grained Pipeline \& Column-based Buffer}}
To ensure lower initial latency and efficient BRAM utilization of the pipeline structure, we adopt the fine-grained layer-based pipeline and column-based cache scheme from \cite{zhang2018dnnbuilder}. With these technologies, we no longer need to wait for the full completion of intermediate results before continuing to the next pipeline stage. Instead, as shown in Figure \ref{fig:dataflow}, the next pipeline stage can be launched once the first few columns or rows of input frame are ready. 
% Figure \ref{fig:pipeline_time}.

% \subsubsection{\textbf{Memory hierarchy? }}

\vspace{-5pt}
\subsection{The Generic Structure}
\vspace{-2pt}
The second part of the proposed architecture is a generic computation unit for processing layer $SP+1 \sim n$. 
% This structure shares similar characteristics with previous DNN accelerators using generic reusable computation units \cite{ye2020hybrid, moreau2019hardware, xilinx_dpu, jouppi2017datacenter}. 
It is composed of an MAC array, a feature maps buffer, a weight buffer, an accumulation buffer, and a functional sub-module for activation and pooling operations. A controller module is included for the data and instructions transfer and management.

\subsubsection{\textbf{Reusable MAC Array}}
The key component of the generic structure is the callable MAC array, which contains $CPF_g \times KPF_g$ units of MAC (implemented with DSPs on FPGAs). The MAC array is able to calculate one general matrix-vector multiplication (GEMV) between a $CPF_g$-length vector of input feature map and a $CPF_g \times KPF_g$ matrix of DNN parameters in each clock cycle. The data width of the on-chip memory bus between the MAC array and weight, feature maps, and accumulation buffer is $CPF_g \times KPF_g \times WW_g$, $CPF_g 
\times DW_g$, and $KPF_g \times DW_g$ bits, respectively, where $WW_g$ and $DW_g$ represents the quantization width of weights and feature maps of the generic structure.

\subsubsection{\textbf{On-chip Buffers Allocation}}
Two types of resource on FPGAs (BRAMs and LUTs) can be utilized to allocate on-chip buffers in the generic structure. BRAMs and LUTs have vastly different characteristics in terms of memory capacity and allocation granularity, which makes them suitable for different usage scenarios. 
% Considering the utilization of BRAMs and LUTs is constrained by the targeted FPGAs, the on-chip buffer allocation should be carefully optimized for generating a balanced and efficient configuration.
%
Two main strategies are found in previous literature as: (1) allocating BRAMs for feature map and accumulation buffer while LUTs for weight buffer (Xilinx DPU \cite{xilinx_dpu}); and (2) allocating BRAMs for all on-chip buffers (VTA \cite{moreau2019hardware} and HybridDNN \cite{ye2020hybrid}). The first strategy allocates most of BRAMs to the feature map buffer, which can effectively reduce the frequency of loading/saving intermediate feature maps from/to the external memory. Meanwhile, the first strategy allows a flexible adjustment for the depth of the feature map buffer to accommodate the BRAMs constraint of the targeted FPGA. The second strategy allocates most of BRAMs to the weight buffer, allowing controller to schedule computations with more freedom, which can potentially benefit the performance but usually consumes more BRAMs than the first strategy. In DNNExplorer, we construct resource and performance models (Section 6) for both strategies to better explore the optimized designs.
%As we discussed above, the data width of on-chip buffers is determined by parallel factors ($CPF_g$ and $KPF_g$) and quantization width ($WW_g$ and $DW_g$).

\subsubsection{\textbf{Feature Map Partitioning}}
Ideally, intermediate feature maps should be fully buffered by on-chip memory for reducing the high-cost data transfer between FPGA and external memory. However, more and more emerging DNN applications are taking high-resolution image or video inputs, demanding a large on-chip buffer capacity, which cannot be accommodated by limited BRAMs on FPGA. To support these emerging applications, we follow a line-based partitioning strategy to break down large feature maps into small groups along the dimension of height. With this strategy, 
the group that completes computations will be swapped out of the on-chip buffer for other new groups. However, such frequent group swapping occupies additional memory access bandwidth, which can impact the accelerator performance. In Section 6, we will introduce how we capture these negative impacts in our modeling algorithms to help deliver optimized designs with better trade-offs.

\vspace{-5pt}
\section{Accelerator Modeling}
\vspace{-2pt}

One of the most critical part in DNNExplorer is the \textit{Accelerator Modeling} that provides fast and accurate estimation of hardware performance and resource utilization. Only with a timely feedback can the decision-making algorithm in \textit{Architecture Exploration} make the most informed decisions.
% These models take the layer-wise information extracted by the \textit{Model/HW Analysis} step and tunable accelerator parameters (e.g. $CPF$, $KPF$, $DW$, and $WW$) as input, and generate estimated performance and FPGA resource utilization, which will serve the optimization in the \textit{Architecture Exploration} step.
Our modeling supports most of commonly-used DNN layers (e.g., CONV, POOL, and FC layers), and can also be extended to support new DNN layers (e.g., depth-wise CONV layers). Considering CONV is the most computation-intensive operation in modern DNNs, in this section, we will take a DNN model with $N$ CONV layers as example to illustrate how we provide the pipeline and generic structure modeling. We assume the $i$-th CONV layer is calculated with a 3-dim input feature $D$ (size $H \times W$ with $C$ channels) and a 4-dim kernel $G$ (size $R \times S$ with $K$ output and $C$ input channels) with clock frequency as $FREQ$.

\vspace{-5pt}
\subsection{Pipeline Structure Modeling}
\vspace{-2pt}

In the pipeline structure, we employ dedicated IPs to carry out the computation of the DNN model following a CTC-based resource allocation algorithm (with detailed discussions in Section 7). The goal of this algorithm is to ensure a load-balanced pipeline structure with perfect match between computation and memory resources for maximal performance. Since a two-dim parallelism factor ($CPF_i$, $KPF_i$) is adopted, the latency $L_i$ of the $i$-th CONV layer is:
\begin{equation}
\small
    L_i = \frac{H_i \times W_i \times R_i \times S_i \times C_i \times K_i}{CPF_i \times KPF_i \times FREQ}
\end{equation}

Since the  throughput maximum is limited by the pipeline stage with the longest latency, we use Eq. \ref{eq:pip_tp} to calculate the overall throughput performance with a batch size of $Batch$.
\begin{equation}
\label{eq:pip_tp}
\small
    Throughput_{p} = \frac{Batch}{max(L_1, L_2, ..., L_N)} 
\end{equation}
%\begin{equation}
%\small
%    P_{model} = \frac{Batch \times \sum_{i=1}^{N} \left ( H_i \times W_i \times R_i \times S_i \times C_i \times K_i \right )}{max(L_1, L_2, ..., L_N)} 
%\end{equation}

\begin{figure}
    %\vspace{-20pt}
    \centering
    \includegraphics[width=0.5\textwidth]{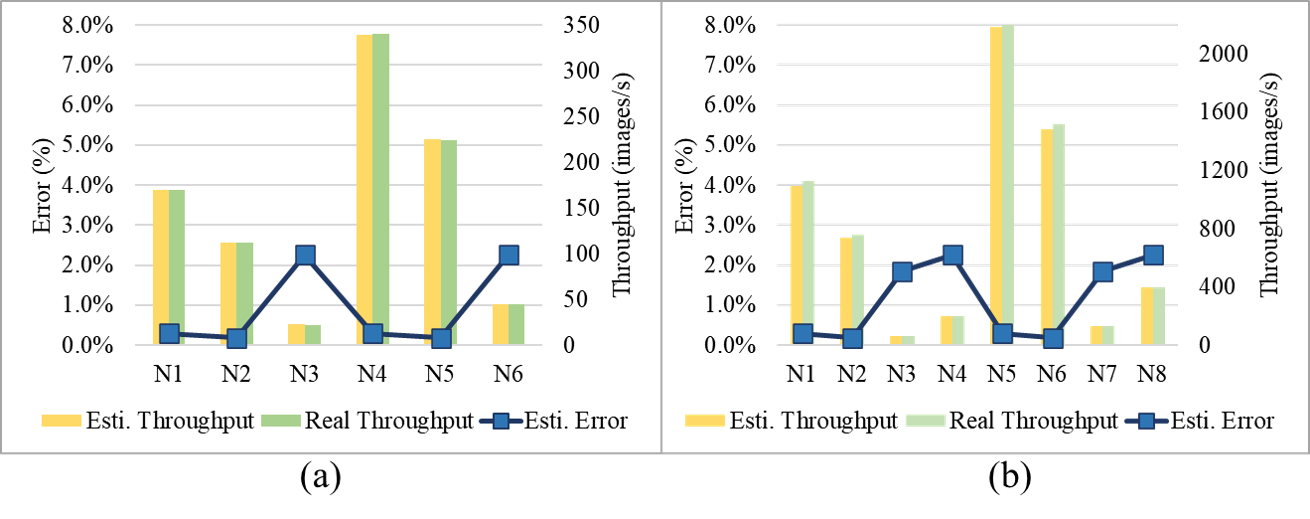}
    \vspace{-20pt}
    \caption{(a) Performance estimation errors of the pipeline structure model when mapping 6 DNNs onto Xilinx ZC706. N1$\sim$N3 represent AlexNet, ZF, and YOLO with 16-bit quantization, while N4$\sim$N6 represent the same group of networks using 8-bit quantization. (b) We swap the FPGA to Xilinx KU115 and evaluate the estimation errors on 8 DNNs where N1$\sim$N4 represent AlexNet, ZF, VGG16, and YOLO with 16-bit quantization while the rest 4 using 8-bit quantization.}
    \label{fig:esti_error_pip}
    \vspace{-10pt}
\end{figure}

%error part
We summarize the estimated throughput performance and the error of the proposed pipeline model in Figure \ref{fig:esti_error_pip}. We evaluate 6 DNN models on an embedded FPGA (Xilinx ZC706) and 8 DNN models on a mid-range FPGA (Xilinx KU115). Results show that the average error between the estimated and the board-level performance is 1.15\%, which verifies the accuracy of our proposed analytical model.

\vspace{-5pt}
\subsection{Generic Structure Modeling}
\vspace{-2pt}

As we discussed in Section 5, two types of on-chip buffer allocation strategies are considered here. For the first strategy, BRAMs are used for keeping feature map and accumulation buffer while for the second strategy, BRAMs are allocated for all on-chip buffers.

\subsubsection{\textbf{Allocation Strategy 1}}
We follow a weight reuse scheme to perform CONV layer. The reuse factor of each weight is determined by the capacity of the accumulation buffer. Assuming the capacity (in bits) of the accumulation buffer is $CAP_{abuff}$, the input and output feature map will be organized as $G_{fm}$ groups where:
\vspace{-2pt}
\begin{equation}
\small
    \label{equ:g_fmap}
    G_{fm} = \frac{H \times W \times K \times DW_g}{CAP_{abuff} / 2} .
\vspace{-1pt}
\end{equation}

In Eq. \ref{equ:g_fmap}, $CAP_{abuff}$ is divided by a factor of 2 because of using ping-pong buffers to avoid data pollution. Pixels of the feature map inside the same group can reuse the same weights which locate on chip. It means the accelerator needs to fetch weights $G_{fm}$ times before finishing all computations in this CONV layer.
The latency of computation $L_{comp}$ and weight loading $L_{w}$ can be calculated as:
\vspace{-2pt}
\begin{equation}
\small
    L_{comp} = \frac{H \times W \times R \times S \times C \times K}{CPF \times KPF \times FREQ}
\end{equation}
\begin{equation}
\small
    \label{equ:l_weight}
    L_{w} = \frac{R \times S \times C \times K \times WW_g}{BW}
\vspace{-1pt}
\end{equation}

In Eq. \ref{equ:l_weight}, $BW$ represents the external memory bandwidth allocated to the generic structure. Then, the overall latency of this CONV layer $L_{layer}$ is:
\vspace{-2pt}
\begin{equation}
\small
    \label{equ:l_all}
    L_{layer} = max(L_{comp}, L_{w} \times G_{fm})
\vspace{-1pt}
\end{equation}

However, due to the feature map partitioning strategy discussed in Section 5.3, the real case is more complicated than Eq. \ref{equ:l_all}.
When the feature map buffer is insufficient, feature maps need to be partitioned and swapped between on-/off-chip memory. To capture the impact of this mechanism, we first divide the external memory bandwidth $BW$ into three portion, $BW_{w}$, $BW_{ifm}$, and $BW_{ofm}$, according to the data access behaviors related to weight loading, feature map swapping in, and feature map swapping out, respectively. Then, the latency $L_{ifm}$ and $L_{ofm}$ can be calculated as:
\vspace{-2pt}
\begin{equation}
\small
    L_{ifm} = \frac{H \times W \times C \times DW_g}{BW_{ifm}}, \quad L_{ofm} = \frac{H \times W \times K \times DW_g}{BW_{ofm}} ,
\vspace{-1pt}
\end{equation}
while the weight loading latency $L_{w}$ and the overall latency $L_{layer}$ are updated as:
\vspace{-6pt}
\begin{equation}
\small
    L_{w} = \frac{R \times S \times C \times K \times WW_g}{BW_{w}}
\end{equation}
\begin{equation}
\small
    \label{equ:l_all_acc1}
    L_{layer} = max(L_{comp}, L_{w} \times G_{fm}, L_{ifm}, L_{ofm})
\vspace{-1pt}
\end{equation}

%To evaluate the accuracy of the proposed model, we use the 15 VGG-16 models shown in Figure \ref{fig:inputsize_ctc} as benchmarks, and compare the estimated performance with the measured performance of an on-board FPGA implementation. The observed error rates (\%) are shown in Figure \ref{fig:esti_error_acc}, and a x.xx\% average error is found across 15 cases of VGG-16 model.

\begin{figure}
    \vspace{-26pt}
    \centering
    \includegraphics[width=0.48\textwidth]{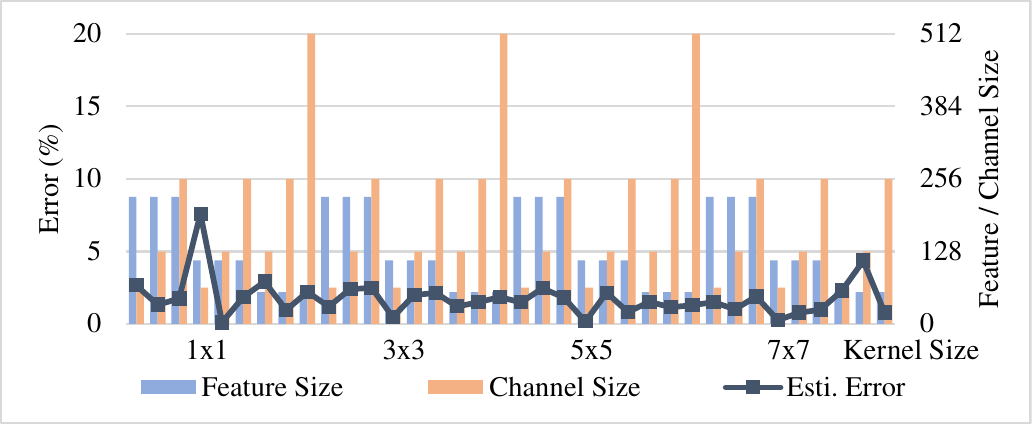}
    \vspace{-20pt}
    \caption{Performance estimation errors of the generic structure when mapping 36 cases of CONV layers. Test cases cover commonly-used feature map sizes (56, 112, 224), channel sizes (64, 128, 256, 512), and kernel sizes (1, 3, 5, 7).}
    \label{fig:esti_error_acc}
    \vspace{-15pt}
\end{figure}

\subsubsection{\textbf{Allocation Strategy 2}}
Similar to strategy 1, to accommodate the capacity of on-chip buffers, we partition input and output feature maps into $G_{fm}$ groups along the dimension of height, and partition weights into $G_{w}$ groups along the dimension of output channels $K$. If we use $CAP_{wbuff}$ to represent the capacity of weight buffer, $G_{w}$ can be calculated as:
\vspace{-1pt}
\begin{equation}
\small
    G_{w} = \frac{R \times S \times C \times K \times WW_g}{CAP_{wbuff}/2}
\vspace{-2pt}
\end{equation}

Under the second allocation strategy, two types of dataflow are supported as input stationary (IS) and weight stationary (WS). For IS, the computation and memory access behaviors are similar to the case using the first strategy, and the overall latency can be calculated by Eq. \ref{equ:l_all_acc1}. However, in this case, because most of BRAMs are allocated to the weight buffer, $G_{fm}$ is usually larger than using the first strategy. It may damage the overall latency according to Eq. \ref{equ:l_all_acc1}. Meanwhile, smaller feature map buffer will prevent the evolution from Eq. \ref{equ:l_all_acc1} to Eq. \ref{equ:l_all}, and increase the external memory bandwidth demands compared to the first strategy.
For WS, the proposed accelerator keeps weights on chip, and for each group of weights, all input feature maps need to be loaded before any computations. Since there are totally $G_{w}$ groups of weights to be calculated, the overall latency is:
\vspace{-2pt}
\begin{equation}
\small
    \label{equ:l_all_acc2}
    L_{layer} = max(L_{comp}, L_{w}, L_{ifm} \times G_{w}, L_{ofm} \times G_{w})
\vspace{-2pt}
\end{equation}

To evaluate the analytical models of the generic structure, we include 36 cases of CONV layers with different channels, feature map, and kernel configurations as benchmarks. We compare the estimated performance with the measured board-level performance using a Xilinx VU9P FPGA. Results are shown in Figure \ref{fig:esti_error_acc}, where only a 2.17\% average error is observed across all 36 cases.
% , which guarantees an effective design space exploration in the \textit{Architecture Exploration} step.
\vspace{-5pt}
\section{Design Space Exploration}
\vspace{-2pt}

\vspace{-1pt}
\subsection{Design Space Definition}
\vspace{-2pt}
With the new accelerator design paradigm, we are allowed to explore a significantly larger design space and perform a more fine-grained hardware resource allocation. We define a dynamic design space regarding all possible accelerator design combinations in Table \ref{tab:design_space}. Split-point defines the task partitioning between the pipeline and generic structure. With more layers distributed to the pipeline structure, more stages are instantiated along with higher dimensions of $CPF_p$ and $KPF_p$.  
Given resource constraints $[DSP, BRAM, BW]$ of the targeted FPGA, DNNExplorer is designed to explore all design parameters and generate the best accelerators for targeted DNN applications. For the efficiency of the exploration, we adopt divide and conquer strategy and propose a two-level automatic DSE engine as the global and local optimization, which will be introduced in the next two subsections.

\begin{table}[t]
\footnotesize
%\scriptsize
\vspace{-20pt}
\caption{Design space of the proposed paradigm}
\vspace{-10pt}
\label{tab:design_space}
\begin{center}
\newcommand{\tabincell}[2]{\begin{tabular}{@{}#1@{}}#2\end{tabular}}
\begin{tabular}{c|c|c}
\toprule
& Pipeline Structure & Generic Structure\\ \hline
\multirow{3}{*}{Parameters}  & \begin{tabular}[c]{@{}c@{}} \tabincell{c}{  $CPF_p = $ \{$CPF_1, CPF_2, \cdots CPF_i$\} \\ $KPF_p = $\{$KPF_1, KPF_2, \cdots KPF_i$\}  }\end{tabular}  & \begin{tabular}[c]{@{}c@{}} \tabincell{c}{  $CPF_g$ \\ $KPF_g$ \\$Buffer$-$allocation$\\$Dataflow$ }\end{tabular}
\\ \cline{2-3}
& \multicolumn{2}{c}{ $Split$-$point (SP)$ } \\ \cline{2-3}
& \multicolumn{2}{c}{ $Batch$ } \\ \hline

%  & $Batch_p$  & \begin{tabular}[c]{@{}c@{}} \tabincell{c}{$Batch_g$\\$Dataflow$}\end{tabular} \\ \hline
Constraints & \multicolumn{2}{c}{ Available $DSP, BRAM, BW$}   \\ 
\bottomrule
\end{tabular}
\vspace{-15pt}
\end{center}
\end{table}

\vspace{-5pt}
\subsection{Global Optimization}
\vspace{-2pt}

Global optimization is the first task of the \textit{Architecture Exploration} step, where updated RAV is generated to provide task partitioning and resource allocation guidelines for our unique accelerator architecture.
To select the best RAV across high-dimensional design space, we employ a particle swarm optimization (PSO) algorithm to discover the most suitable combination of $SP$, $Batch$, and hardware resource distribution between the pipeline and generic structure. The global optimization can be also extended to support other optimization algorithms in the future for different scenarios.

As shown in Algorithm \ref{alg:global}, each RAV is considered as a particle $P_i$, and all of them contribute to the swarm $\mathcal{M}$. We use throughput performance as the fitness index to evaluate the performance of each particle. We label the local best for particle $i$ across all iterations as $L_i$ and use $G$ to represent the global best particle. The search contains  $\mathcal{N}$ iterations and in the $itr$-iteration, the position of each particle is updated according to the current position and updated velocity $V_i$. The updated velocity is calculated based on the velocity to the local $V_{toLbest_i}$ and global $V_{toGbest_i}$ best position. In the update function, $rand()$ 
generates random numbers between 0 to 1 and we also include the adjustable parameters as inertia weight, $w$, acceleration constants, $c_1$ and $c_2$ to fine-tune the search process. Eventually, the best particle is selected, which is the best RAV indicating the optimal task partitioning and resource allocation. For improving the search efficiency of the global optimization, we introduce an early termination feature, which will terminate the optimization if $G$ is not improved for two continuous iterations.

\floatstyle{spaceruled}% Select new float style
\restylefloat{algorithm}% Apply spaceruled float style to algorithm
\begin{algorithm}[t!]
\renewcommand\baselinestretch{0.5}\selectfont
\footnotesize
\caption{Global optimization algorithm}
\begin{algorithmic}[1]
\State Initialize RAV with $\mathcal{M}$ Population: $P(\mathcal{M})$
\State Initialize iteration number: $\mathcal{N}$
\State Initialize HW boundary: $SP_{max}$, $DSP_{max}$, $BRAM_{max}$, $BW_{max}$
\State Evaluate each RAV: $Fit_i= FitnessScore(P_i)$, where $i \in \mathcal{M}$
\State Keep the local best for each RAV: $L_i = Fit_i$, and the global best: $G=max(L_i)$
\State \textbf{While} $itr < \mathcal{N}$:
\State \quad \textbf{For} each $P_i$ in $\mathcal{M}$:
\State \quad \quad Get local and global velocity: $V_{toLbest_i}$, $V_{toGbest_i}$ 
\State \quad \quad Update velocity: $V_i = w \cdot V_i + c_1 \cdot rand() \cdot V_{toLbest_i} + c_2 \cdot rand() \cdot V_{toGbest_i} $
\State \quad \quad Update RAV: $P_i = UpdatePos(P_i,V_i)$
\State \quad \quad Evaluate updated RAV: $Fit_i = FitnessScore(P_i) $
\State \quad \quad Update local best: $L_i$
% \State \quad \textbf{EndFor}
\State \quad Update global best: $G$
% \State \textbf{EndWhile}
\State Output the best RAV
\end{algorithmic}
\label{alg:global}
\end{algorithm}

\vspace{-5pt}
\subsection{Local Optimization}
\vspace{-2pt}

Once RAV is updated by the global optimizer, the local optimization will be launched to search for the best parameters (e.g., $CPF_p$, $KPF_p$, $CPF_g$, $KPF_g$, and $Batch$) combination given the constraint of RAV. Considering the flexible architecture of our proposed accelerator enables a huge design space, a brute-force search algorithm indicates an exponentially increasing complexity and is infeasible to find the optimal solution. To solve this problem, we propose a two-phase local optimization algorithm as shown in Algorithm \ref{alg:local_pip} (phase 1) and Algorithm \ref{alg:local_gen} (phase 2). In the first phase, we will calculate $CPF$ and $KPF$ of each layer in the pipeline structure based on workload and CTC characteristics, and ensure the computation resource and external memory bandwidth is utilized to the maximum. In the second phase, we increase $CPF_g$ and $KPF_g$ step by step until the performance of the pipeline and the generic structure is balanced, which means the overall latency of the generic structure is less or equal to the max latency of the pipeline structure. 
If the FPGA resources are exhausted under the balanced point, we will roll back and scale down $CPF$s and $KPF$s of the pipeline structure, until all resource constraints are met.
In the second phase, Algorithm \ref{alg:local_gen} will be executed for each on-chip buffer allocation strategy discussed in subsection 5.3.2, and the best one will become the final solution. If the second strategy is selected, the latency update in line 9 of Algorithm \ref{alg:local_gen} will automatically select the better dataflow configuration (IS or WS) for each layer.

\begin{algorithm}[t]
\renewcommand\baselinestretch{0.8}\selectfont
\footnotesize
\caption{CTC-based local optimization algorithm for the pipeline structure}
\begin{algorithmic}[1]
\State $[SP, Batch, DSP_p, BRAM_p, BW_p] = RAV$
\State Initialize latency and resource model, $lpip()$ and $rpip()$, for pipeline structure.
\State Calculate operation number $OP_i$ and computation reuse factor $CTC_i$ for each layer of the targeted DNN model.
\State $BW_{total}^{norm} = \sum_{i=1}^{SP} ( OP_i / CTC_i )$
\State \textbf{for} $i$ \textbf{in} [$1, SP$] :
\State \quad $PF_i = \lceil OP_i \times BW_p / BW_{total}^{norm} / FREQ \rceil$
\State \textbf{while} ($\sum_{i=1}^{SP} R_i \geq [DSP_p, BRAM_p]$) :
\State \quad \textbf{for} $i$ \textbf{in} [$1, SP$] :
\State \quad \quad $PF_i = max(1, PF_i / 2)$, update $CPF_i$ and $KPF_i$.
\State \quad \quad Update $L_i$ and $R_i$ with model $lpip()$ and $rpip()$.
\end{algorithmic}
\label{alg:local_pip}
% \vskip -0.1in
\end{algorithm}

\begin{algorithm}[t]
\renewcommand\baselinestretch{0.8}\selectfont
\footnotesize
\caption{Balance-oriented local optimization algorithm for the generic structure}
\begin{algorithmic}[1]
\State Initialize latency and resource model, $lgen()$ and $rgen()$, for generic structure.
\State Initialize $R_{total}$ with the total resource boundary of the targeted FPGA.
\State \textbf{while} (1) :
\State \quad $PF_{g}=1$, $L_{p}^{max} = max(L_1, L_2, ..., L_{SP})$
\State \quad \textbf{while} ($\sum_{i = SP+1}^{N} L_i > L_{p}^{max}$ \textbf{and} $R_{g} + \sum_{i=1}^{SP} R_i < R_{total}$) :
\State \quad \quad $PF_{g} = PF_{g} \times 2$, update $CPF_{g}$ and $KPF_{g}$.
\State \quad \quad Update $R_{g}$ with resource model $rgen()$.
\State \quad \quad \textbf{for} i \textbf{in} [$SP+1, N$] :
\State \quad \quad \quad Update $L_i$ with latency model $lgen()$.
\State \quad $Batch_{p} = (R_{total} - R_{g}) / \sum_{i=1}^{SP} R_i$
\State \quad \textbf{if} ($\sum_{i = SP+1}^{N} L_i > L_{p}^{max}$ \textbf{or} $Batch > Batch_{p}$) :
\State \quad \quad \textbf{for} i \textbf{in} [$1, SP$] :
\State \quad \quad \quad $PF_i = max(1, PF_i / 2)$, update $CPF_i$ and $KPF_i$.
\State \quad \quad \quad Update $L_i$ and $R_i$ with model $lpip()$ and $rpip()$.
\State \quad \textbf{else} : \textbf{break}
\end{algorithmic}
\label{alg:local_gen}
% \vskip -0.1in
\end{algorithm}
\begin{table*}[t]
\footnotesize
%\scriptsize
\vspace{-20pt}
\caption{Performance and resource overhead of the DNNExplorer-generated accelerators with batch size = 1.}
\vspace{-10pt}
\label{tab:rav_gops}
\begin{center}
\newcommand{\tabincell}[2]{\begin{tabular}{@{}#1@{}}#2\end{tabular}}
\begin{tabular}{c|c|c|c|c|c|c|c|c}
\toprule
Case  & Input Size & GOP/s & Img./s & $R =$ [$SP, DSP, BRAM, BW$] & Total DSP & DSP Efficiency & Total BRAM & Avg. Search Time (s)\\ \hline
1 & 3x32x32  & 368.5 & 588.9 & [4, 48.2\%,	17.7\%,	62.9\%] & 2268 & 42.3\% & 2326 & 86.6
\\ \hline
2 & 3x64x64  & 890.8 & 339.1 & [5, 46.5\%,	13.5\%,	65.9\%] & 2730 & 77.9\% & 2560 & 41.6\\ \hline
3 & 3x128x128  & 1702.3 & 169.5 & [9, 50.2\%,	38.9\%,	54.2\%] & 4686 & 90.8\% & 3589 & 70.2\\ \hline
4 & 3x224x224  & 1702.3 & 55.4 & [12, 63.6\%,	53.7\%,	67.3\%] & 4444 & 95.8\% & 3296 & 75.6\\ \hline
5 & 3x320x320  & 1702.4 & 27.1 & [13, 73.5\%,	71.4\%,	63.5\%] & 4450 & 95.7\% & 3224 & 45.9\\ \hline
6 & 3x384x384  & 1702.4 & 18.8 & [14, 73.2\%,	70.1\%,	55.0\%] & 4452 & 95.6\% & 3436 & 64.9\\ \hline
7 & 3x320x480  & 1702.4 & 18.1 & [14, 78.6\%,	74.6\%,	58.2\%] & 4452 & 95.6\% & 3296 & 71.1\\ \hline
8 & 3x448x448  & 1702.4 & 13.8 &[13, 75.8\%,	72.9\%,	36.1\%] & 4450 & 95.6\% & 3552 & 82.1\\ \hline
9 & 3x512x512  & 1702.4 & 10.6 &[13, 80.0\%,	70.0\%,	80.0\%] & 4450 & 95.6\% & 3678 &70.0\\ \hline
10 & 3x480x800  & 1702.4 & 7.2 &[13, 80.0\%,	72.0\%,	80.0\%] & 4450 & 95.6\% &  3678 & 103.0\\ \hline
11 & 3x512x1382  & 1702.5 & 3.9 &[14, 77.4\%,	73.1\%,	34.3\%] & 4452 & 95.6\% & 3792 & 105.5\\ \hline
12 & 3x720x1280  & 1702.5 & 3.0 &[13, 82.7\%,	73.4\%,	84.0\%] & 4450 & 95.6\% & 4186 & 143.9\\ 

\bottomrule
\end{tabular}
\vspace{-10pt}
\end{center}
\end{table*}

\vspace{-5pt}
\section{Experimental Results}
\vspace{-2pt}

% In this section, we demonstrate the great capability of  DNNExplorer to deliver FPGA-based DNN accelerator designs with improved performance, higher efficiency, and much better scalability compared to the state-of-the-art solutions. 

% \vspace{-5pt}
\subsection{DSP Efficiency Comparison}
\vspace{-2pt}

\begin{figure}
    \vspace{-0pt}
    \centering
    \includegraphics[width=0.36\textwidth]{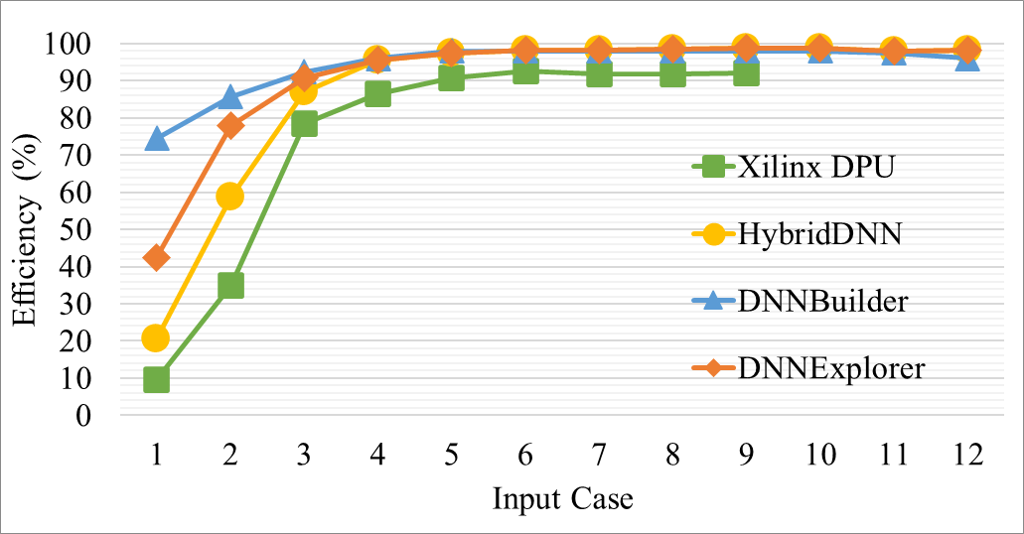}
    \vspace{-10pt}
    \caption{DSP efficiency comparison when running VGG16 (batch size = 1) with 12 different input sizes.}
    \label{fig:inputVSeff}
    \vspace{-10pt}
\end{figure}

To accelerate DNN applications using FPGAs, DSP is the most important and  scarce computation resources. In this subsection, we use DSP efficiency as criteria to evaluate whether accelerators maximize the usage of allocated DSPs. We compare our design to two other automatic accelerator design frameworks, DNNBuilder \cite{zhang2018dnnbuilder} and HybridDNN \cite{ye2020hybrid}, by targeting the same task: to deliver DNN accelerators on a Xilinx KU115 FPGA corresponding to 12 VGG-16 (without the last three FC layers) models with different input sizes (as the same cases in Figure \ref{fig:inputsize_ctc}) to simulate tasks of real-life DNN applications. In our experiments, weights and feature maps are quantized to 16-bits fixed-point in all three frameworks for equality. Results are shown in Figure \ref{fig:inputVSeff}, where DNNBuilder achieves the highest DSP efficiency especially targeting small size inputs (e.g, case 1 and 2) because of its dedicated pipeline stage design. Our proposed design is slightly behind DNNBuilder when targeting small inputs but we then reach the same efficiency level (>95\%) after case 3. Compared to the generic accelerator design in HybridDNN, our design can deliver 2.0x and 1.3x higher efficiency for case 1 and 2, respectively.
We also compare to Xilinx DPU \cite{xilinx_dpu}, a commercial DNN accelerator IP, for running first 9 cases on a ZCU102 FPGA (as the last three inputs are not supported by this IP). The accelerators generated by DNNExplorer can achieve an average 1.6x higher DSP efficiency, peaking at 4.4x for case 1.  
As the input size increases, the efficiency gap decreases (<10\% after case 5).

\begin{figure}
    \centering
    \includegraphics[width=0.4\textwidth]{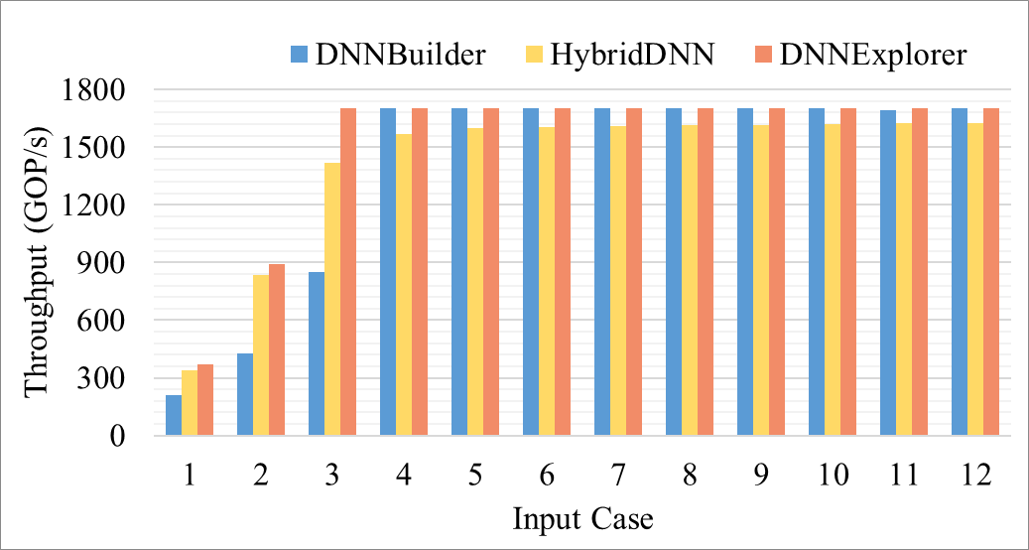}
    \vspace{-10pt}
    \caption{Throughput comparison when running VGG16 (batch size = 1) with 12 different input sizes.}
    \label{fig:inputVSgops}
    \vspace{-10pt}
\end{figure}

\begin{figure}
 \vspace{-0pt}
    \centering
    \includegraphics[width=0.38\textwidth]{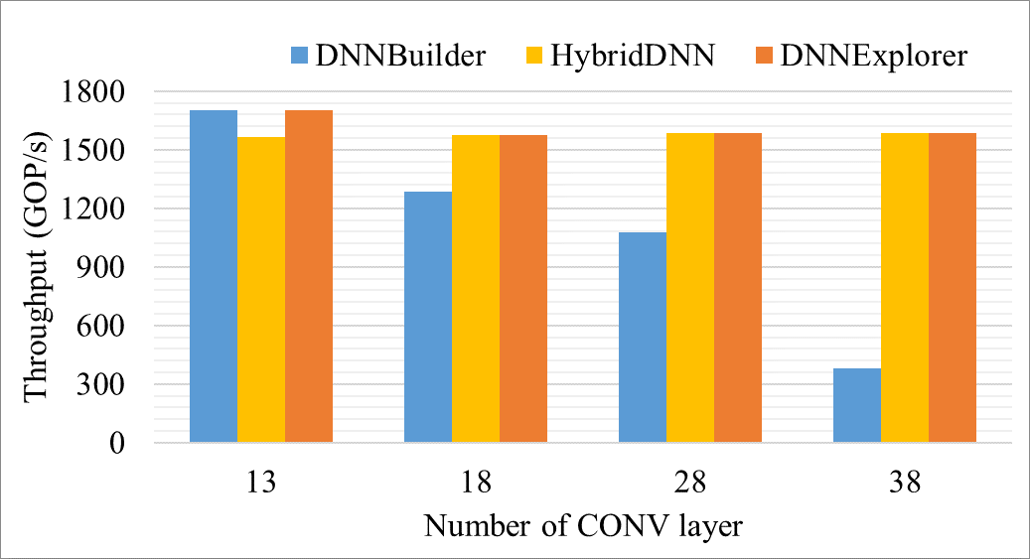}
    \vspace{-10pt}
    \caption{Throughput comparison when running deeper DNNs with the same $3\times224\times224$ input.}
    \label{fig:depthVSgops}
    \vspace{-15pt}
\end{figure}

\vspace{-5pt}
\subsection{Performance Comparison}
\vspace{-2pt}

We compare the throughput (GOP/s) among accelerators generated by the proposed DNNExplorer, DNNBuilder, and HybridDNN. First, we target the same DNNs mentioned in the last subsection using Xilinx KU115 FPGA with 200MHz clock frequency. Performance results are listed in Figure \ref{fig:inputVSgops}, where accelerators generated by DNNExplorer achieve better throughput performance compared to the state-of-the art solutions. 
To meet resource constraints and maximize
% the external memory access efficiency, the first on-chip buffer allocation strategy is selected by the DSE for the generic structure of the produced design.
performance, the proposed DSE engine generates configuration guidelines for implementing the proposed architecture.

Detailed results are shown in Table \ref{tab:rav_gops}, where $R$ indicates the most suitable task partition scheme and resources allocated to the $P$ and $G$ structures of the proposed accelerator. The total DSP and total BRAM represent the compute and on-chip memory utilization. To capture the runtime of searching the optimal architecture given input model and hardware constraints, we perform five independent searches using Intel i5-650 CPU for each case and list the average search time. With the divide-and-conquer idea and the early termination feature, DNNExplorer delivers the best configurations in minutes regardless the high-dimensional design space. 
We extend the search for batch size in Table \ref{tab:dse_w_batch}, where lower DNN complexity (caused by smaller inputs in case 1 to 4) creates opportunities for batch size exploration given limited available FPGA resources. 

% For simplicity, we only present the \textit{SP} and the percentage of resources  allocated to the first structure, while rest of resources belong to the second structure. 
% With larger input size, DNNExplorer intends to instantiate more dedicated pipeline stages to maintain high DSP efficiency (e.g., case \#1 $\sim$ \#9). As the inputs become larger, DNNExplorer reduces the pipeline stages to avoid overuse of the computation and memory resources and shift the focus to building bigger generic structures.

We also evaluate the scalability of DNNExplorer by targeting  deeper DNNs. We prepare 4 DNNs, with 13, 18, 28, and 38 CONV layers. The 13-layer one comes directly from VGG16 by removing the last 3 FC layers. Since VGG is composed of 5 CONV groups, where each group has the same CONV configurations (e.g, number of CONV kernels), we add one CONV layer to each group (maintaining the same configurations) and get the 18-layer (13+5) model. Similarly, we add 3 and 5 CONV layers to each part for the 28- and 38-layer model, respectively. In Figure \ref{fig:depthVSgops}, our designs maintain the highest performance despite targeting deeper networks. Compared to DNNBuilder, we can deliver 4.2x higher performance for accelerating a 38-layer VGG-like DNN on the same FPGA platform.

\begin{table}[t]
\footnotesize
%\scriptsize
% \vspace{-15pt}
\caption{Performance and resource overhead of the generated accelerators without batch size restrict.}
\vspace{-10pt}
\label{tab:dse_w_batch}
\begin{center}
\newcommand{\tabincell}[2]{\begin{tabular}{@{}#1@{}}#2\end{tabular}}
\begin{tabular}{c|c|c|c|c|c|c}
\toprule
Case  & Input Size & Batch & GOP/s & Img./s  & DSP & BRAM \\ \hline
1 & 3x32x32 & 8 & 1698.1 & 2712.7 & 4464 &  2228 \\ \hline
2 & 3x64x64  & 8 & 1701.5 & 678.2  & 4688  & 3534 \\ \hline
3 & 3x128x128 & 4 & 1702.4 & 169.5  & 4686  & 3326 \\ \hline
4 & 3x224x224 & 2 & 1702.3 & 55.4  & 4686  & 3619 \\

\bottomrule
\end{tabular}
\vspace{-15pt}
\end{center}
\end{table}
\vspace{-5pt}
\section{Conclusion}
\vspace{-2pt}

\quad In this paper, we presented a novel FPGA-based DNN accelerator design paradigm with both dedicated pipeline structure and generic reusable structure to overcome the drawbacks of existing designs by providing improved performance, better specificity, and scalability. To achieve the full potential of the new paradigm, we proposed DNNExplorer, an automation tool to enable fast architecture exploration following the new paradigm. Novel technologies were proposed which included a dynamic design space for more fine-grained architecture configurations and a two-level automatic DSE engine for delivering optimized designs. With the above novel designs, DNNExplorer achieved 4.2x higher performance compared to the state-of-the-art design produced from DNNBuilder when targeting a 38-layer VGG-like DNN on the same FPGA. We also achieved higher DSP efficiency (up to 4.4x)  compared to the latest generic architecture accelerators from academia (HybridDNN) and industry (Xilinx DPU).

\section*{Acknowledgment}
\vspace{-2pt}
\quad This work was supported in part by the IBM-Illinois Center for Cognitive Computing System Research (C$^3$SR) -- a research collaboration as part of IBM AI Horizons Network. Xiaofan Zhang is supported by a Google PhD Fellowship.

\vspace{-8pt}

\renewcommand*{\bibfont}{\footnotesize}

\bibliographystyle{unsrt}
\bibliography{sample-bibliography} 

\begin{thebibliography}{10}

\bibitem{zhang2018dnnbuilder}
Xiaofan Zhang et~al.
\newblock {DNNBuilder: an automated tool for building high-performance DNN
  hardware accelerators for FPGAs}.
\newblock In {\em Proceedings of International Conference on Computer-Aided
  Design (ICCAD)}, 2018.

\bibitem{ye2020hybrid}
Hanchen Ye et~al.
\newblock {HybridDNN}: A framework for high-performance hybrid {DNN}
  accelerator design and implementation.
\newblock In {\em Proceedings of the Design Automation Conference (DAC)}, 2020.

\bibitem{xilinx_dpu}
Xilinx.
\newblock Zynq dpu v3.1, 2019.
\newblock Accessed: 2020-5-23.

\bibitem{krizhevsky2012imagenet}
Alex Krizhevsky, Ilya Sutskever, and Geoffrey~E Hinton.
\newblock Imagenet classification with deep convolutional neural networks.
\newblock In {\em Advances in neural information processing systems}, 2012.

\bibitem{simonyan2014very}
Karen Simonyan and Andrew Zisserman.
\newblock Very deep convolutional networks for large-scale image recognition.
\newblock {\em arXiv preprint arXiv:1409.1556}, 2014.

\bibitem{szegedy2015going}
Christian Szegedy et~al.
\newblock Going deeper with convolutions.
\newblock In {\em Proceedings of the IEEE conference on computer vision and
  pattern recognition (CVPR)}, 2015.

\bibitem{he2016deep}
Kaiming He et~al.
\newblock Deep residual learning for image recognition.
\newblock In {\em Proceedings of the IEEE conference on computer vision and
  pattern recognition (CVPR)}, 2016.

\bibitem{redmon2016you}
Joseph Redmon et~al.
\newblock You only look once: Unified, real-time object detection.
\newblock In {\em Proceedings of the IEEE conference on computer vision and
  pattern recognition (CVPR)}, 2016.

\bibitem{zoph2018learning}
Barret Zoph et~al.
\newblock Learning transferable architectures for scalable image recognition.
\newblock In {\em Proceedings of the IEEE conference on computer vision and
  pattern recognition (CVPR)}, 2018.

\bibitem{real2019regularized}
Esteban Real et~al.
\newblock Regularized evolution for image classifier architecture search.
\newblock In {\em AAAI conference on Artificial Intelligence (AAAI)}, 2019.

\bibitem{zhang2020skynet}
Xiaofan Zhang et~al.
\newblock {SkyNet}: a hardware-efficient method for object detection and
  tracking on embedded systems.
\newblock In {\em Conference on Machine Learning and Systems (MLSys)}, 2020.

\bibitem{franklin2018nvidia}
Dustin Franklin.
\newblock {NVIDIA Jetson AGX Xavier} delivers 32 teraops for new era of {AI} in
  robotics.
\newblock {\em NVIDIA Accelerated Computing| Parallel For all}, 2018.

\bibitem{jouppi2017datacenter}
Norman~P Jouppi et~al.
\newblock In-datacenter performance analysis of a tensor processing unit.
\newblock In {\em Proceedings of International Symposium on Computer
  Architecture (ISCA)}, 2017.

\bibitem{isscc_2016_chen_eyeriss}
Yu-Hsin Chen et~al.
\newblock {Eyeriss}: An energy-efficient reconfigurable accelerator for deep
  convolutional neural networks.
\newblock In {\em IEEE International Solid-State Circuits Conference (ISSCC)},
  2016.

\bibitem{zhang2015optimizing}
Chen Zhang et~al.
\newblock Optimizing {FPGA}-based accelerator design for deep convolutional
  neural networks.
\newblock In {\em Proceedings of the International Symposium on
  Field-Programmable Gate Arrays (FPGA)}, 2015.

\bibitem{xu2020autodnnchip}
Pengfei Xu et~al.
\newblock {AutoDNNchip}: An automated {DNN} chip predictor and builder for both
  {FPGAs} and {ASICs}.
\newblock 2020.

\bibitem{qiu2016going}
Jiantao Qiu et~al.
\newblock {Going deeper with embedded FPGA platform for convolutional neural
  network}.
\newblock In {\em Proceedings of International Symposium on Field-Programmable
  Gate Arrays (FPGA)}, 2016.

\bibitem{zhang2017high}
Xiaofan Zhang et~al.
\newblock High-performance video content recognition with long-term recurrent
  convolutional network for {FPGA}.
\newblock In {\em Proceedings of the International Conference on Field
  Programmable Logic and Applications (FPL)}, 2017.

\bibitem{zhang2017machine}
Xiaofan Zhang et~al.
\newblock Machine learning on {FPGAs} to face the {IoT} revolution.
\newblock In {\em Proceedings of International Conference on Computer-Aided
  Design (ICCAD)}, 2017.

\bibitem{zhang2017improving}
Jialiang Zhang and Jing Li.
\newblock Improving the performance of opencl-based fpga accelerator for
  convolutional neural network.
\newblock In {\em Proceedings of the International Symposium on
  Field-Programmable Gate Arrays (FPGA)}, 2017.

\bibitem{hao2019fpga}
Cong Hao et~al.
\newblock {FPGA/DNN} co-design: An efficient design methodology for {IoT}
  intelligence on the edge.
\newblock In {\em Proceedings of the Design Automation Conference (DAC)}, 2019.

\bibitem{li2016high}
Huimin Li et~al.
\newblock A high performance fpga-based accelerator for large-scale
  convolutional neural networks.
\newblock In {\em Proceedings of the International Conference on Field
  Programmable Logic and Applications (FPL)}, 2016.

\bibitem{wei2018tgpa}
Xuechao Wei et~al.
\newblock {TGPA}: tile-grained pipeline architecture for low latency cnn
  inference.
\newblock In {\em Proceedings of the International Conference on Computer-Aided
  Design (ICCAD)}, 2018.

\bibitem{zhao2017accelerating}
Ritchie Zhao et~al.
\newblock Accelerating binarized convolutional neural networks with
  software-programmable {FPGAs}.
\newblock In {\em Proceedings of the International Symposium on
  Field-Programmable Gate Arrays (FPGA)}, 2017.

\bibitem{wang2018design}
Junsong Wang et~al.
\newblock Design flow of accelerating hybrid extremely low bit-width neural
  network in embedded {FPGA}.
\newblock In {\em Proceedings of the International Conference on Field
  Programmable Logic and Applications (FPL)}, 2018.

\bibitem{xiao2017exploring}
Qingcheng Xiao et~al.
\newblock Exploring heterogeneous algorithms for accelerating deep
  convolutional neural networks on fpgas.
\newblock In {\em Proceedings of the Design Automation Conference (DAC)}, 2017.

\bibitem{zhuge2018face}
Chuanhao Zhuge et~al.
\newblock Face recognition with hybrid efficient convolution algorithms on
  {FPGAs}.
\newblock In {\em Proceedings of the Great Lakes Symposium on VLSI (GLSVLSI)},
  2018.

\bibitem{han2017ese}
Song Han et~al.
\newblock {ESE}: Efficient speech recognition engine with sparse lstm on
  {FPGA}.
\newblock In {\em Proceedings of the International Symposium on
  Field-Programmable Gate Arrays (FPGA)}, 2017.

\bibitem{zhang2018caffeine}
Chen Zhang et~al.
\newblock Caffeine: Toward uniformed representation and acceleration for deep
  convolutional neural networks.
\newblock {\em IEEE Transactions on Computer-Aided Design of Integrated
  Circuits and Systems}, 38(11):2072--2085, 2018.

\bibitem{wei2017automated}
X.~Wei et~al.
\newblock Automated systolic array architecture synthesis for high throughput
  {CNN} inference on {FPGAs}.
\newblock In {\em DAC}, 2017.

\bibitem{rupnow2011study}
Kyle Rupnow et~al.
\newblock A study of high-level synthesis: Promises and challenges.
\newblock In {\em IEEE International Conference on ASIC}, 2011.

\bibitem{liu2016high}
Xinheng Liu et~al.
\newblock High level synthesis of complex applications: An h. 264 video
  decoder.
\newblock In {\em Proceedings of International Symposium on Field-Programmable
  Gate Arrays (FPGA)}, 2016.

\bibitem{guan2017fp}
Yijin Guan et~al.
\newblock Fp-dnn: An automated framework for mapping deep neural networks onto
  {FPGAs} with {RTL-HLS} hybrid templates.
\newblock In {\em Proceedings of the International Symposium on
  Field-Programmable Custom Computing Machines (FCCM)}, 2017.

\bibitem{li2019implementing}
Qin Li et~al.
\newblock Implementing neural machine translation with bi-directional {GRU} and
  attention mechanism on {FPGAs} using {HLS}.
\newblock In {\em Proceedings of Asia and South Pacific Design Automation
  Conference (ASP-DAC)}, 2019.

\bibitem{chen2019cloud}
Yao Chen et~al.
\newblock {Cloud-DNN}: An open framework for mapping {DNN} models to cloud
  {FPGAs}.
\newblock In {\em Proceedings of the International Symposium on
  Field-Programmable Gate Arrays (FPGA)}, 2019.

\bibitem{moreau2019hardware}
Thierry Moreau, Tianqi Chen, Luis Vega, Jared Roesch, Eddie Yan, Lianmin Zheng,
  Josh Fromm, Ziheng Jiang, Luis Ceze, Carlos Guestrin, et~al.
\newblock A hardware--software blueprint for flexible deep learning
  specialization.
\newblock {\em IEEE Micro}, 39(5):8--16, 2019.

\end{thebibliography}

\end{document}